\documentclass[
twocolumn,
 superscriptaddress,
 amsmath,amssymb,
 aps,
 prl,
]{revtex4-2}

\usepackage{amsmath}
\usepackage{graphicx}
\usepackage{dcolumn}
\usepackage{bm}
\usepackage{amsfonts}
\usepackage{mathrsfs}
\usepackage{amssymb}
\usepackage{slashed}
\usepackage[colorlinks=true,linkcolor=blue]{hyperref}
\usepackage[capitalize]{cleveref}
\usepackage[x11names]{xcolor}
\usepackage{bm}
\usepackage{empheq}
\usepackage{ragged2e}
\usepackage[normalem]{ulem}
\usepackage{array} 
\usepackage{multirow}

\crefname{section}{Sec.}{Sec.}
\Crefname{section}{Section}{Sections}

\newcommand{\prlsection}[1]{%
\textit{#1}---%
}

\usepackage[super]{nth}

\def\bfk{\mathbf{k}}
\def\bfp{\mathbf{p}}

\usepackage{dsfont}
\usepackage{url}
\usepackage{lipsum}
\usepackage{color}
\usepackage{colortbl}

\begin{document}

\title{Jet Momentum Broadening in Viscous QCD Matter: A Moment Expansion Approach}

\author{Isabella Danhoni}
\email{idanhoni@illinois.edu}
\affiliation{Illinois Center for Advanced Studies of the Universe, Department of Physics, University of Illinois at
Urbana-Champaign, Urbana, IL 61801, USA}
\author{Nicki Mullins}
\email{nmmulli2@ncsu.edu}
\affiliation{Department of Physics, North Carolina State University, Raleigh, NC 27695, USA}
\author{Jorge Noronha}
\affiliation{Illinois Center for Advanced Studies of the Universe, Department of Physics, University of Illinois at
Urbana-Champaign, Urbana, IL 61801, USA}

\begin{abstract}
We formulate out-of-equilibrium jet momentum broadening in QCD effective kinetic theory through a moment expansion of the medium distribution function, a method traditionally used to derive relativistic viscous hydrodynamics from kinetic theory. We explicitly compute the leading near-equilibrium contribution to the spatial jet broadening tensor $\hat q^{ij}$ within the 14-moment approximation, and show that it is controlled by the medium shear-stress tensor. This provides a direct map from QCD effective kinetic theory to event-by-event viscous hydrodynamic simulations, converting local shear-stress fields into anisotropic corrections to jet broadening in heavy-ion collisions.
\end{abstract}

\maketitle


\prlsection{Introduction}
In heavy-ion collisions, jets are collimated sprays of hadrons \cite{Connors:2017ptx} associated with highly energetic partons produced in the initial hard scattering
\cite{Gyulassy:1990ye,Baier:2000mf,Gyulassy:2003mc,Qin:2015srf,Mehtar-Tani:2013pia,Wiedemann:2009sh}. As they propagate through the medium produced in the collision, these partons interact with the surrounding matter and are therefore sensitive to the entire space--time evolution of the quark--gluon plasma (QGP), including the early stages of the collision \cite{Kurkela:2014tea,Qin:2015srf,Ipp:2020nfu,Andres:2019eus,Andres:2022ndd}, where the medium can be far from equilibrium \cite{Schenke:2012wb,Niemi:2014wta,Noronha-Hostler:2015coa}. Because jets propagate over distances comparable to the size of the system, they provide a unique probe of the microscopic dynamics of the evolving QGP \cite{Cao:2020wlm,Mehtar-Tani:2025rty}.

The medium-induced broadening of a high-energy parton is commonly characterized by the jet-quenching parameter $\hat q$. For thermally equilibrated plasmas, this quantity has been computed in a wide range of theoretical frameworks, including weakly coupled QCD
\cite{Baier:2000mf,Aurenche:2002pd,Arnold:2008vd,Caron-Huot:2008zna,Benzke:2012sz,DEramo:2012uzl,Kurkela:2018vqr,Ghiglieri:2015zma,Schlichting:2020lef,Boguslavski:2023waw,Boguslavski:2023alu,Barata:2025agq},
holographic approaches
\cite{Liu:2006ug,Casalderrey-Solana:2011dxg,Grefa:2022sav,Khan:2026uqc}, and other methods
\cite{DEramo:2010wup,Laine:2013apa,Panero:2013pla,Kumar:2020wvb}. Phenomenological estimates of the jet-quenching parameter from model-to-data comparisons have also been carried out by the JET \cite{Burke:2013yra,Xu:2014tda} and JETSCAPE \cite{JETSCAPE:2021ehl} collaborations. More generally, jet momentum broadening can be described by a tensor $\hat q^{ij}$ \cite{Salgado:2003gb,Caron-Huot:2008zna,Ghiglieri:2018dib,Boguslavski:2023alu,Boguslavski:2023waw}, which measures the rate
at which momentum fluctuations are accumulated along the jet trajectory.

In an isotropic and homogeneous equilibrated plasma, rotational symmetry reduces the
spatial momentum-broadening tensor to two scalar functions, a transverse coefficient $\hat q_\perp$ and a longitudinal coefficient $\hat q_\parallel$ \cite{Boguslavski:2023fdm,Barata:2024xwy,Barata:2025agq, Danhoni:2026gve}. However, the medium produced in heavy-ion collisions is neither isotropic nor homogeneous and is generally out of thermal equilibrium. In fact, hydrodynamic simulations indicate that sizable momentum anisotropies and viscous stresses can persist over a significant fraction of the QGP evolution \cite{Romatschke:2017ejr}. Consequently, the symmetry arguments that enforce a diagonal momentum-broadening tensor no longer apply, and one expects the viscous hydrodynamic medium to induce nontrivial anisotropic contributions to $\hat q^{ij}$, especially at early times after the collision.

These considerations have prompted studies of jet--medium interactions
in non-equilibrium settings, including the early glasma phase
\cite{Aurenche:2012qk,Carrington:2022bnv,Lappi:2006fp,Ipp:2020nfu}
and anisotropic or inhomogeneous plasmas
\cite{Barata:2025wnp,Barata:2025htx,Antiporda:2021hpk,Barata:2023zqg,Chernicoff:2012gu,Barata:2025agq,Kuzmin:2023hko,Barata:2022krd}.
In particular, QCD kinetic theory \cite{Arnold:2002zm} has been widely used to study momentum
broadening in non-equilibrium systems
\cite{Schlichting:2020lef,Boguslavski:2023waw,Boguslavski:2023alu,Barata:2025agq,Boguslavski:2024jwr}.
In these works, the medium is modeled by non-equilibrium distribution
functions designed to capture plasma anisotropy. While such approaches
capture important features of out-of-equilibrium jet--medium
interactions, a direct connection between jet momentum broadening and the dissipative currents that characterize the viscous evolution of the relativistically expanding QGP in hydrodynamic simulations has remained elusive. This missing link is particularly important for phenomenology, since
hydrodynamic simulations provide the local dissipative fields of the
medium, such as the shear-stress tensor $\pi^{\mu\nu}(x)$, rather than ad hoc anisotropy
parameters.

In this Letter, we address this problem by expanding in moments the
non-equilibrium correction to the medium distribution function,
following a method widely used to derive relativistic viscous
hydrodynamics from kinetic theory
\cite{Denicol:2012cn,Rocha:2023ilf}. This yields a controlled framework
in which non-equilibrium corrections to jet momentum broadening
generated by the medium are expressed directly in terms of viscous
hydrodynamic variables obtained from numerical simulations. Within the
standard 14-moment approximation
\cite{Israel:1979wp,Denicol:2012cn,Denicol:2012es}, we show that the
leading non-equilibrium correction to \( \hat q^{ij} \) is controlled
by \( \pi^{\mu\nu} \), establishing a straightforward connection
between microscopic jet momentum broadening in QCD and macroscopic
viscous fluid dynamics in the QGP. Since \( \pi^{\mu\nu} \) is a
standard output of viscous hydrodynamic simulations, our result
provides a new direct way to incorporate \emph{event-by-event}
shear-induced corrections into phenomenological studies of jet momentum
broadening in heavy-ion collisions.

\bigskip

\noindent
\emph{Notation:} We use the $(-,+,+,+)$ metric signature
and natural units $\hbar = c = k_{B} = 1$. Momentum 4-vectors are denoted by $K^\mu = (k^0,\bfk)$, and $U^\mu$ is the medium's 4-velocity.  

\bigskip

\prlsection{Jet momentum broadening in QCD kinetic theory} We treat the jet as a tagged, dilute probe with momentum $P^\mu$  propagating through a medium described by single-particle
distribution functions \cite{Boguslavski:2023alu}. The collision operator is used only to extract the broadening tensor governing the probe, without solving for the backreaction of the jet perturbation on the medium. In the high-momentum probe limit, jet occupation factors are neglected, while medium
statistical factors are retained \cite{Boguslavski:2023fdm}. This separation is particularly well motivated in QCD where momentum broadening is dominated by multiple soft scatterings mediated by gluon exchange, leading to a parametrically controlled small-angle diffusion limit \cite{Arnold:2002zm,Arnold:2003zc,Ghiglieri:2015zma,Ghiglieri:2015ala,Ghiglieri:2018dib}. Thus, in this regime, jet momentum broadening is generated by the cumulative effect of microscopic scatterings with medium constituents \cite{Caron-Huot:2008zna,Danhoni:2026gve,Barata:2024xwy} and $\hat q^{ij}$ can be expressed as
\begin{equation}
\begin{split}
\hat q^{ij}
&=
\frac{(2\pi)^4}{4 p\, \nu_a}
\sum_{bcd}
\int d\Gamma_{k p' k'} \delta^{(4)}(P+K-P'-K')\\
&\times \left| \mathcal{M}^{ab}_{cd} \right|^2 \, q^i q^j \,
  f_b(X,K)\,
\tilde f_c(X,P')\,
\tilde f_d(X,K') \,,
\end{split}
\label{eq:qhat_def}
\end{equation}
where $Q^\mu = P'^\mu - P^\mu$ is the 4-momentum transfer, 
$p = |\mathbf{p}|$ is the magnitude of the jet's 3-momentum,
$\nu_a$ denotes the color degeneracy of the jet parton, $f_{b}(X,K)$ denotes the distribution function for particle species $b$, i.e., either a quark or a gluon (also, note that $\tilde f_{b}=1\pm f_{b}$ with the upper/lower
sign corresponding to bosons/fermions). Above, the Lorentz-invariant phase-space measure is denoted by,
\begin{equation}
d\Gamma_{ k  p'  k'} \equiv 
\frac{d^3\bfk}{(2\pi)^3 k^0}
\frac{d^3\mathbf p'}{(2\pi)^3 p'^0}
\frac{d^3\bfk'}{(2\pi)^3 k'^0}\, .
\end{equation}
The squared matrix element 
$\left| \mathcal{M}^{ab}_{cd} \right|^2$ is taken to be the standard QCD 
scattering matrix element computed within hard thermal loop (HTL) approximation~\cite{Arnold:2003zc,Kurkela:2018oqw,Schlichting:2020lef}. Following the $\xi$-screening
prescription in kinetic-theory calculations introduced in
\cite{AbraaoYork:2014hbk}, and derived for transverse momentum diffusion in the context of jet momentum broadening in~\cite{Boguslavski:2024jwr}, the matrix element is given by
\begin{equation}
\mathcal{M}_{cd} ^{ab}\rightarrow \mathcal{M}_{\xi}
=
\frac{s - u}{t^2}
\,
\frac{q^4}{\left(q^2 + \xi^2 m_D^2\right)^2},
\end{equation}
where $m_D^2 = N_c/3g^2T^2+N_F/6g^2(T^2+3/\pi^2\mu^2)$ is the Debye mass squared \cite{Rischke:2003mt}, and $s$, $u$, and $t$ are the usual Mandelstam variables \cite{Peskin:1995ev}.
For asymptotically energetic jets, $p\to\infty$, momentum broadening in QCD is dominated 
by small-angle $t$-channel processes, which are logarithmically enhanced, as 
discussed in~\cite{Arnold:2000dr,Boguslavski:2023waw}.

To incorporate the corrections from equilibrium to jet momentum broadening in a systematic fashion, we expand $f_b(X,K)$ around local equilibrium as,
\begin{align}
\nonumber
f_b(X,K) &= f_{0,b}(X,K) + \delta f_b(X,K)\\
&= f_{0,b}(X,K)\left[1+\tilde{f}_{0,b}(X,K)\phi_k\right],
\label{eq:distr_func}
\end{align}
where $f_{0,b}(X,K) =
1/\left(\exp\!\left[-\beta\,E_\bfk\right] \mp 1\right)$ denotes the local equilibrium distribution, $E_\bfk = -U_\mu K^\mu$, and $\beta = 1/T$ is the inverse temperature. The scalar function $\phi_k$ thus encodes all the non-equilibrium contributions coming from the medium. 
This implies that one can write $\hat q^{ij}
=
\hat q^{ij}_{\rm eq}
+
\delta \hat q^{ij}$, where the equilibrium piece is
\begin{eqnarray}
\hat q^{ij}_{\mathrm{eq}}
&=&
\frac{1}{4 p\, \nu_a}
\sum_{bcd}
\int d\Gamma_{k p' k'}
\,
\left| \mathcal{M}^{ab}_{cd} \right|^2
q^i q^j \\ \nonumber &\times&
(2\pi)^4
\delta^{(4)}(P+K-P'-K')
f^b_{0k}\tilde f^d_{0k'} ,
\label{eq:define_equilibrium_qij}
\end{eqnarray}
and, for small deviations from equilibrium, $\delta \hat q^{ij}$ collects all contributions that are \emph{linear} in out-of-equilibrium deviations 
$\delta f$, which we write as
\begin{align}
\nonumber
\delta \hat q^{ij}&
=
\frac{(2\pi)^4}{4 p\, \nu_a}
\sum_{bcd}
\int d\Gamma_{k p' k'}\delta^{(4)}(P+K-P'-K')
\,
\\
&\times q^i q^j \left| \mathcal{M}^{ab}_{cd} \right|^2
\left[
f^b_{0k} f^d_{0k'} \tilde f^d_{0k'} \phi_{k'} + f^b_{0k} \tilde f^b_{0k}  \tilde f^d_{0k'} \phi_k
\right].
\label{eq:define_delta_qij}
\end{align}

The equilibrium contribution has been extensively studied before, see \cite{Boguslavski:2023alu,Boguslavski:2023waw,Ghiglieri:2015ala,Caron-Huot:2008zna,Ghiglieri:2015zma} (see also our discussion about $\hat q^{ij}_{\mathrm{eq}}$ below). In this paper, following \cite{Denicol:2012cn}, we go beyond previous works by expanding $\phi_k$ in a \textit{complete and orthogonal} basis formed by the irreducible tensors constructed using $K^\mu$, $U^\mu$, and the metric $g_{\mu\nu}$
\begin{equation}
1,\; K^{\langle \mu \rangle},\; K^{\langle \mu}K^{\nu\rangle},\;
K^{\langle \mu}K^{\nu}K^{\lambda\rangle},\ldots .
\end{equation}
Angle brackets denote symmetric, traceless projections orthogonal to the
fluid 4-velocity, 
\begin{equation}
A^{\langle\mu\rangle} = \Delta^\mu_{\;\nu}A^\nu,
\qquad
\Delta_{\mu\nu}=g_{\mu\nu}+U_\mu U_\nu ,
\end{equation}
with higher-rank projections defined analogously.
The expansion then reads \cite{Denicol:2012cn}
\begin{align}
\phi_k =
\sum_{\ell=0}^{\infty}\sum_{n=0}^{N_\ell}
\mathcal H_{\mathbf k n}^{(\ell)}\,
\rho_n^{\mu_1\cdots\mu_\ell}
K_{\langle\mu_1}\cdots K_{\mu_\ell\rangle},
\end{align}
where the moments are defined as
\begin{equation}
\rho_n^{\mu_1\cdots\mu_\ell}(X) = \int \frac{d^3\bfk}{(2\pi)^3 k^0}  E_\bfk^n\, K^{\langle\mu_1}\cdots K^{\mu_\ell\rangle}\,\delta f(X,K),
\end{equation}
with $\mathcal H_{\mathbf k n}^{(\ell)}$ being polynomials that depend only
on $E_\bfk$, and 
the integer $\ell$ labels the rank of the irreducible tensors \cite{Denicol:2012cn}. Typical matching conditions for the hydrodynamic fields (Landau matching \cite{Rocha:2023ilf}) correspond to setting $\rho_1=\rho_2=\rho_1^\mu=0$. In this notation, for example, the shear-stress tensor is $\pi^{\mu\nu} = \rho_0^{\mu\nu}$ \cite{Denicol:2012cn}.  

Exact equations of motion for the moments
$\rho_n^{\mu_1\cdots\mu_\ell}$ can be obtained from the underlying
integro-differential Boltzmann equation \cite{Denicol:2012cn}. In
principle, solving this infinite hierarchy of coupled partial
differential equations is equivalent to solving the Boltzmann equation
itself and can be used to reconstruct the full distribution function;
see \cite{Bazow:2015dha,Bazow:2016oky,Mullins:2022fbx} for explicit
examples in relativistic systems. For the present work, however,
Eq.~\eqref{eq:define_delta_qij} shows that any approximation for
$\phi_k$ obtained either from the moment hierarchy itself or from a
physically motivated truncation scheme yields an approximation
to the out-of-equilibrium contribution $\delta \hat q^{ij}$ generated by
the medium. For simplicity, in this paper we adopt the lowest-order truncation
compatible with causal and stable hydrodynamic evolution
\cite{Rocha:2023ilf}, namely the 14-moment approximation
\cite{IsraelStewart1979,UdeyIsrael1982}. The resulting expression for
$\delta \hat q^{ij}$ can later be systematically improved by including
additional moments \cite{Denicol:2012cn}. 


\prlsection{14-moment approximation}
All moments contribute in principle to the full microscopic expression in \eqref{eq:define_delta_qij}. However, higher-rank moments are associated with increasingly fine structures in momentum space and are
expected to relax on microscopic time scales. Near equilibrium, one can therefore truncate the expansion by retaining only the lowest-rank moments and neglecting higher-rank contributions. In four dimensions, the 14-moment approximation is obtained by setting irreducible moments with $\ell\ge 3$ to zero and expressing the remaining scalar, vector, and rank-two moments in terms of the dissipative currents. This yields a total of 14 independent degrees of freedom~\cite{IsraelStewart1979,UdeyIsrael1982}.
The retained moments of the distribution function then describe, in addition to the energy density, particle/baryon density, and four-velocity already present in ideal hydrodynamics, the dissipative
corrections to the conserved currents, namely bulk pressure, particle diffusion, and shear stress. This is precisely the standard moment closure used to derive relativistic hydrodynamics from kinetic theory
\cite{IsraelStewart1979,Denicol:2012cn,Denicol:2012es}.

Here, we consider massless particles at zero baryon chemical potential.
We work in the Landau frame, so that the energy-diffusion current
vanishes by our choice of matching conditions. In addition, at zero baryon chemical potential the charge-diffusion current $\mathcal{J}^{\mu}$ vanishes,
and for our massless system the bulk viscous scalar $\Pi=0$ (we neglect running-coupling
corrections that generate a nonzero bulk viscosity coefficient
\cite{Arnold:2006fz}). The only remaining dissipative contribution is
therefore the shear-stress tensor $\pi^{\mu\nu}$. Consequently, the
nonequilibrium correction $\phi_k$ can depend only on $\pi^{\mu\nu}$ and takes the form 
\begin{align}
\phi_k = \frac{1}{2J_{42}} \, \pi_{\mu\nu} K^\mu K^\nu .
\label{eq:connects_deltaq_ij_and_pi_ij}
\end{align}
The coefficient $J_{42}$ is a standard temperature-dependent thermodynamic integral
appearing in the 14-moment approximation~\cite{Denicol:2012cn} (see the Supplemental Material). Thus, using \eqref{eq:define_delta_qij} and \eqref{eq:connects_deltaq_ij_and_pi_ij}, the non-equilibrium corrections to jet momentum broadening generated by the medium can be written directly in terms of $\pi^{\mu\nu}$, establishing the desired connection with viscous hydrodynamics. 


\prlsection{Computation of the jet momentum broadening tensor}
We perform the calculation of $\hat q^{ij}$ in the fluid local rest frame (LRF) where
$U^\mu=(1,\mathbf{0})$, the shear tensor is purely spatial
$\pi^{0\mu}=0$ and $\phi_k =  \, \pi_{ij} k^i k^j/2J_{42}$ (with $\pi^{ij}\delta_{ij}=0$). We choose $\hat{n}$ to define the jet momentum so $P^\mu = (E_\bfp,p \,\hat{n})$. 
Due to rotational symmetry, the equilibrium jet momentum broadening tensor \eqref{eq:define_equilibrium_qij} in the LRF must take the form
\begin{equation}
\hat q^{ij}_{\rm eq}
=
\hat q^{\rm eq}_\perp
\left(\delta^{ij}-\hat n^i \hat n^j\right)
+
\hat q^{\rm eq}_\parallel
\,\hat n^i \hat n^j\, .
\label{eq:qhat_eq_LRF}
\end{equation}
Rotational symmetry around the jet axis therefore restricts the
equilibrium tensor to two independent scalar quantities, $\hat q_\perp^{\rm eq}$ and $\hat q_\parallel^{\rm eq}$, which encode the transverse and longitudinal momentum broadening of the jet in equilibrium, respectively. Within the HTL approximation, we show in the Supplemental Material that we recover $\hat q^{\rm eq}_\perp = \frac{g^2}{4\pi}C_ATm_D^2\ln{\left(1+\frac{\Lambda_\perp^2}{m_D^2}\right)}$, and $\hat q^{\rm eq}_\parallel = \frac{g^2 C_A T\, M_\infty^2}{4\pi}
\ln\!\left(1 + \frac{\mu_{\tilde q_\perp}^2}{M_\infty^2}\right)$ (where $M_\infty^2=\frac{g^2 C_A T^2}{6}$), which have been previously computed in the literature \cite{Ghiglieri:2015zma} ( $\Lambda_\perp$ denotes the upper bound for the momentum exchanged between the jet and the medium parton).  

Away from equilibrium, $\pi^{ij}$ introduces additional anisotropic behavior for the jet broadening tensor. Working in linear response around local equilibrium, \eqref{eq:define_delta_qij} implies that  
\begin{equation}
\delta \hat q^{ij}
=
\mathcal{K}^{ijab}\,
\pi_{ab}.
\label{eq:deltaqhat_linear}
\end{equation}
The response tensor 
$\mathcal{K}^{ijab}$ is fixed by the collision kernel and inherits its dependence on the underlying QCD scattering matrix elements. Furthermore, this quantity is constrained by rotational symmetry in the local rest frame, as well as by symmetry under $i \leftrightarrow j$ and $a \leftrightarrow b$. It can therefore be constructed from the available tensors $\delta^{ij}$ and the jet direction $\hat n^i$ and its most general form in the LRF is given by
\begin{align}
\nonumber
\delta \hat q^{ij}
&=
\alpha\,\pi^{ij}
+
\beta\,\delta^{ij}\,\hat n^a \hat n^b \pi_{ab}\\
&+
\gamma\,
\left(
\hat n^i \pi^{j a}\hat n_a
+
\hat n^j \pi^{i a}\hat n_a
+
\hat n^i \hat n^j \hat n^a \hat n^b \pi_{ab}
\right)
\label{eq:deltaqijfull}
\end{align}
where the Lorentz scalar coefficients $\alpha$, $\beta$, and $\gamma$ encode the microscopic dynamics and depend directly on the
jet--medium interactions. We remark that Eq.~\eqref{eq:deltaqijfull} follows solely from rotational symmetry
in the local rest frame and linearity in the shear-stress tensor. Thus,
while the scalar coefficients $\alpha$, $\beta$, and $\gamma$ are
computed here in QCD effective kinetic theory, the tensorial structure itself is model independent and therefore must arise in any framework describing the linear nonequilibrium response of jet momentum broadening to $\pi^{ij}$; only the values of the scalar coefficients are framework dependent.

The explicit expressions for the coefficients $\alpha$, $\beta$, and $\gamma$ in QCD effective kinetic theory are (see the Supplemental Material for a detailed derivation): 
\begin{equation}
   \begin{split}
\alpha
=&\,\frac{g^4}{4\nu_a(2\pi)^5}
\int d\Gamma_{ k  k'p'}\;\delta^{(4)}(P+K-P'-K')
\tilde{\mathcal{M}}_{\rm screen}^2\\
&\times q^x q^y f_{0k}\tilde f_{0k'}\left[
\frac{f_{0k'}}{2J_{42}}\,k'^xk'^y +\frac{\tilde f_{0k}}{2J_{42}}\,k^xk^y \right],
\label{eq:alpha}
  \end{split} 
\end{equation}
\begin{equation}
   \begin{split}
    \beta
=&\,\frac{g^4}{4\nu_a(2\pi)^5}
\int d\Gamma_{ k  k'p'}\;\delta^{(4)}(P+K-P'-K')
\tilde{\mathcal{M}}_{\rm screen}^2
\\
&\times q^x q^x f_{0k}\tilde f_{0k'}
\left[
\frac{f_{0k'}}{2J_{42}}\,k'^zk'^z
+
\frac{ \tilde f_{0k}}{2J_{42}}\,k^zk^z
\right],
\label{eq:beta}
 \end{split} 
\end{equation}
\begin{equation}
    \begin{split}
\gamma
=&\,\frac{g^4}{4\nu_a(2\pi)^5}
\int d\Gamma_{ k   k'p'}\;\delta^{(4)}(P+K-P'-K')
\tilde{\mathcal{M}}_{\rm screen}^2\\
&\times q^x q^z f_{0k}\tilde f_{0k'}
\left[
\frac{f_{0k'}}{2J_{42}}\,k'^xk'^z
+
\frac{ \tilde f_{0k}}{2J_{42}}\,k^xk^z
\right]
-\alpha,
    \end{split}
    \label{eq:gamma}
\end{equation}  
where $\tilde{\mathcal{M}}_\mathrm{screen}^2=\lim_{p\to\infty} \mathcal{M}_{\xi}^2/g^4p^2$ as in~\cite{Boguslavski:2023waw}. 

The three independent tensor structures admit a clear physical interpretation. The first coefficient, $\alpha$, corresponds to
the \textit{direct shear imprint channel} describing a direct transfer of the medium anisotropy into the jet-broadening tensor, aligning the principal axes of $\hat q^{ij}$ with those of $\pi^{ij}$.
The second coefficient, $\beta$, modifies the overall
magnitude of spatial momentum broadening in proportion to the projection
of the shear-stress tensor along the jet direction (e.g., the $z$ direction), $\delta \hat q^{ij} \sim
\delta^{ij}\pi^{zz}$,
which may be interpreted as an \textit{isotropic spatial shift} of the
broadening strength in the fluid rest frame.  
Finally, the \textit{jet-direction--selective}
structure associated with $\gamma$, $\delta\hat q^{ij} \sim \hat n^i\hat n^j\pi^{zz}+\hat n^i\pi^{jz} + \hat n^j\pi^{iz}$,
modifies momentum broadening along the jet direction relative to
transverse directions in momentum space. This contribution is governed
by the projection of the shear tensor onto the jet momentum.
Together, these structures exhaust the independent tensor contributions
linear in $\pi^{ij}$ that can appear in
$\delta \hat q^{ij}$ in the LRF. 

To better illustrate how these contributions work, consider a local rest frame configuration with jet
direction $\hat n=\hat z$ and a diagonal shear tensor
$\pi^{ij}=\mathrm{diag}(-\pi_s/2,-\pi_s/2,\pi_s)$, with $\pi_s>0$. In this case,
$\delta \hat q^{xx} = \left(-\frac{\alpha}{2}+\beta\right)\pi_s$,
$\delta \hat q^{yy} = \left(-\frac{\alpha}{2}+\beta\right)\pi_s$, and
$\delta \hat q^{zz} = (\alpha+\beta+3\gamma)\pi_s$,
while off-diagonal components vanish. Thus, even for this simplest shear
pattern, viscous corrections split longitudinal and transverse momentum
broadening in a way that depends on the distinct microscopic channels
encoded in $\alpha$, $\beta$, and $\gamma$.

\begin{figure}
    \centering
    \includegraphics[width=\linewidth]{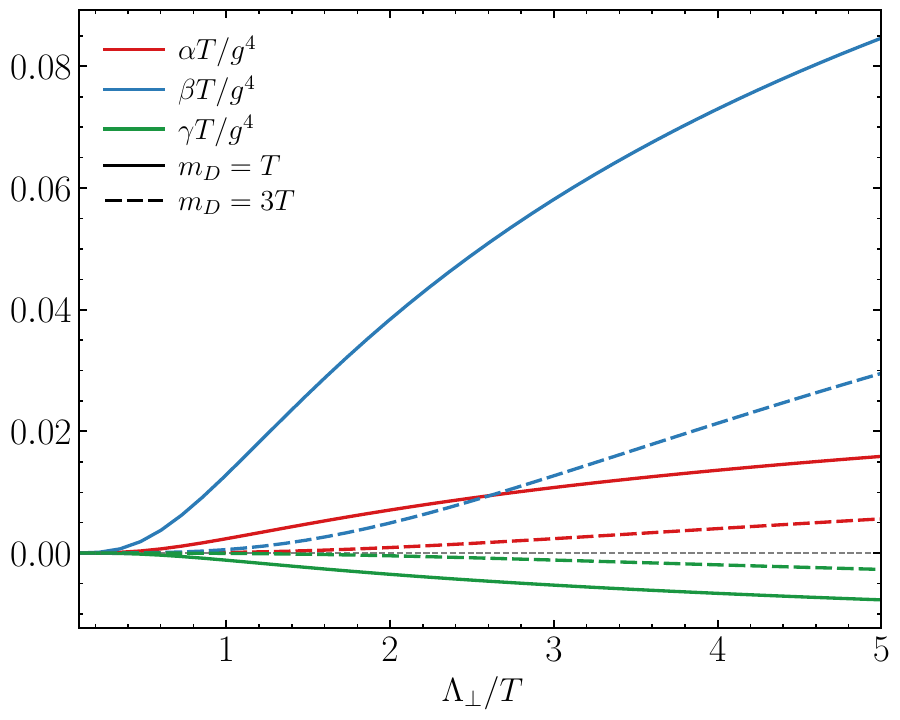}
    \caption{Coefficients $\alpha T$, $\beta T$, and $\gamma T$ defined in 
Eqs.~\eqref{eq:alpha}, \eqref{eq:beta}, and \eqref{eq:gamma} as a function 
of the transverse momentum cutoff $\Lambda_\perp/T$ for a purely gluonic 
medium, computed with screened $t$-channel gluon exchange. Solid and dashed 
lines correspond to $m_D = T$ and $m_D = 3T$, respectively.}
    \label{fig:plot}
\end{figure}

Figure~\ref{fig:plot} shows the coefficients $\alpha$, $\beta$, and
$\gamma$ as functions of $\Lambda_\perp$, for two
values of the screening mass in a purely gluonic medium with screened
$t$-channel exchange. The same formalism applies to quark scattering,
with the corresponding changes in statistical factors, color
degeneracies, and scattering matrix elements. We find that increasing $m_D$
suppresses the magnitude of all three coefficients over the range shown, consistent with the
physical expectation that stronger screening reduces momentum
broadening, in agreement with equilibrium calculations
\cite{Boguslavski:2023waw}. We observe that $\alpha$, $\beta$, and $\gamma$ preserve the
equilibrium growth with the momentum cutoff, with the diagonal contribution $\beta$ exhibiting a substantially larger magnitude than the off-diagonal coefficients. This hierarchy is expected, since the diagonal projection maximizes the contraction
$q_i^2 \cdot k_i^2$, whereas the off-diagonal components do not share this structure. From the plot, the off-diagonal coefficients appear to show a weaker dependence on the momentum cutoff than the diagonal contribution, in qualitative agreement with equilibrium expectations.
For the cases
shown in Fig.~\ref{fig:plot}, we find $\alpha>0$, $\beta>0$, and
$\gamma<0$. The positivity of $\beta$ is expected, since it involves the
manifestly positive combination $(q^x k^z)^2$, while the negative sign
of $\gamma$ reflects the nontrivial angular structure of the
corresponding projection, as detailed in the Supplemental Material.
This shows that shear-induced corrections are not sign-definite and can
enhance or suppress different components of the broadening tensor.
Therefore, the full tensorial structure of $\delta \hat q^{ij}$ must be
retained for a consistent description of jet broadening out of
equilibrium.

Beyond the shear sector studied here, the same framework can in
principle be extended to include bulk viscous pressure and
charge-diffusion currents, which would generate additional tensor
structures in $\delta \hat q^{ij}$. Furthermore, following arguments from \cite{Dusling:2009df},  one could also investigate the case where momentum anisotropy is carried by medium particles with somewhat lower momenta than in the 14-moment Ansatz, which might affect the response coefficients $\alpha$, $\beta$, and $\gamma$.

Finally, we remark that a fully covariant extension of the present framework is also possible,
following the general considerations of Ref.~\cite{Danhoni:2026gve}. In
such a formulation, the momentum-broadening coefficient is promoted to a
rank-two tensor $\delta \hat q^{\mu\nu}$ depending on the jet momentum,
the local fluid 4-velocity, and the dissipative currents. In the
local rest frame, its spatial components reduce to
\eqref{eq:deltaqijfull}. The explicit covariant construction of $\delta \hat q^{\mu\nu}$ is left
for future work.

\prlsection{Conclusions}
In this Letter, we developed a systematic framework to compute
out-of-equilibrium jet momentum broadening in QCD effective kinetic
theory by expanding the medium distribution function in moments. 
To obtain a tractable description, we specialized to the standard
14-moment approximation. For a massless system at vanishing baryon
chemical potential, the leading non-equilibrium correction to the
broadening tensor is then controlled by the shear-stress tensor
$\pi^{\mu\nu}$. We derived the corresponding tensor decomposition of
$\delta \hat q^{ij}$ and computed the coefficients $\alpha$, $\beta$,
and $\gamma$ that encode the underlying microscopic QCD dynamics.
Importantly, the tensor decomposition itself is fixed by general
symmetry arguments and is therefore not specific to kinetic theory; QCD effective kinetic theory is used here to evaluate the corresponding microscopic coefficients, alternative approaches remain applicable and may offer complementary descriptions of the underlying microscopic dynamics.

In practice, our results establish a direct connection between microscopic jet
momentum broadening and macroscopic viscous fluid dynamics, and provide
a practical framework for incorporating viscous corrections from
hydrodynamic simulations into jet broadening phenomenology. In fact, since
$\pi^{\mu\nu}(x)$ is a standard output of viscous hydrodynamics, the present framework provides a direct map from the
dissipative contributions from the medium to the anisotropic jet broadening
tensor, making it possible to quantify how shear anisotropies modify jet
broadening in event by event hydrodynamic simulations. Extensions including the effects from higher moments, as well as to a
fully covariant formulation, are left for future work.

\section*{Acknowledgments}
We thank G.~D.~Moore for providing comments on this manuscript. I.D. and J.N. were
partly supported by the US-DOE Nuclear Science Grant No. DE-SC0023861 and by
the National Science Foundation (NSF) within the framework of the MUSES collaboration, under grant number
OAC-2103680.
N.M. was partially
supported by U.S. Department of Energy, Office of Nuclear Physics, Contract DE-FG02-03ER41260.
Any opinions, findings, and conclusions
or recommendations expressed in this material are those
of the author(s) and do not necessarily reflect the views of
the National Science Foundation or U.S. Department of Energy.

\bibliography{main-4,scalar_qhat_v3}

\clearpage

\onecolumngrid

\begin{center}
\textbf{\large Supplemental Material}
\end{center}

In this Supplementary Material, we present details of the analytical calculation of the covariant jet momentum–broadening tensor in equilibrium and its leading non-equilibrium correction within the 14-moment approximation.

\bigskip

\section{Definition of jet momentum broadening}
We begin from the kinetic-theory definition of the jet momentum-broadening tensor,
\begin{equation}
\begin{split}
\hat q^{ij}(P,u)
&=
\frac{(2\pi)^4}{4 p\, \nu_a}
\sum_{bcd}
\int d\Gamma_{k p' k'} \delta^{(4)}(P+K-P'-K') \left| \mathcal{M}^{ab}_{cd} \right|^2 \, q^i q^j \,
f_b(X,K)\,
\tilde f_c(X,P')\,
\tilde f_d(X,K') \,,
\end{split}
\label{eq:qhat_def}
\end{equation}
where $Q^\mu = P'^\mu - P^\mu$ is the four-momentum transfer,
$p = |\mathbf{p}|$ is the magnitude of the jet's three-momentum,
$\nu_a$ denotes the color degeneracy of the jet parton, and $f_b(X,K)$
denotes the distribution function for particle species $b$, with
$\tilde f_b = 1 \pm f_b$ (upper/lower sign for bosons/fermions), where $d$ denotes final-state medium parton, and $c$ denotes the final-state of the jet parton.
The Lorentz-invariant phase-space measure is
\begin{equation}
d\Gamma_{ k  p'  k'} \equiv
\frac{d^3\mathbf{k}}{(2\pi)^3 k^0}
\frac{d^3\mathbf{p}'}{(2\pi)^3 p'^0}
\frac{d^3\mathbf{k}'}{(2\pi)^3 k'^0}\,.
\end{equation}
Here the sum runs over species $b$ and  $d$ for medium particle species . The collision operator is evaluated in the probe limit: jet occupation factors are neglected while medium
statistical factors are retained \cite{Boguslavski:2023fdm}, so that
$\tilde{f}_c(X,P') \approx 1$ in the high-momentum limit $p \to \infty$.

To incorporate deviations from local equilibrium in a controlled way, we expand
$f_b(X,K)$ around local equilibrium as
\begin{align}
\nonumber
f_b(X,K) &= f_{0,b}(X,K) + \delta f_b(X,K)\\
&= f_{0,b}(X,K)\left[1+\tilde{f}_{0,b}(X,K)\phi_k\right],
\label{eq:distr_func}
\end{align}
where $f_{0,b}(X,K) = 1/\!\left(\exp\!\left[-\beta\,E_{\mathbf{k}}\right] \mp 1\right)$
denotes the local equilibrium distribution, $E_{\mathbf{k}} = -U_\mu K^\mu$,
and $\beta = 1/T$ is the inverse temperature. The scalar function $\phi_k$
encodes all non-equilibrium contributions from the medium.

This implies that $\hat q^{ij} = \hat q^{ij}_{\mathrm{eq}} + \delta \hat q^{ij}$,
where the equilibrium piece is
\begin{eqnarray}
\hat q^{ij}_{\mathrm{eq}}
&=&
\frac{1}{4 p\, \nu_a}
\sum_{bd}
\int d\Gamma_{k p' k'}
\,
\left| \mathcal{M}^{ab}_{cd} \right|^2
q^i q^j (2\pi)^4
\delta^{(4)}(P+K-P'-K')
f^b_{0k}\tilde f^d_{0k'} ,
\label{eq:define_equilibrium_qij}
\end{eqnarray}
and, for small deviations from equilibrium, $\delta \hat q^{ij}$ collects all
contributions \emph{linear} in out-of-equilibrium deviations $\delta f$,
\begin{align}
\delta \hat q^{ij}
&=
\frac{(2\pi)^4}{4 p\, \nu_a}
\sum_{bd}
\int d\Gamma_{k p' k'}\delta^{(4)}(P+K-P'-K')
\, q^i q^j \left| \mathcal{M}^{ab}_{cd} \right|^2
\left[
f^b_{0k} f^d_{0k'} \tilde f^d_{0k'} \phi_{k'} + f^b_{0k} \tilde f^b_{0k} \tilde f^d_{0k'} \phi_k
\right].
\label{eq:define_delta_qij}
\end{align}
Higher-order terms in $\phi$ are neglected, retaining only linear deviations from
equilibrium as appropriate within the fourteen-moment approximation and for systems
close to local thermal equilibrium \cite{Arnold:2000dr}. The contribution
proportional to $\phi_{p'}$ has been dropped: since $p'$ labels the final state
of the jet parton after scattering, perturbing this state would amount to modifying
the probe itself rather than the medium, which lies outside the scope of the present
calculation \cite{Boguslavski:2023fdm}.

\section{Jet momentum broadening for an equilibrated medium}
\label{sec:equilibrium}
In this section, we evaluate the equilibrium contribution to the jet
momentum-broadening tensor defined in Eq.~(\ref{eq:qhat_def}). We work in the
local rest frame of the medium and consider the high-energy limit of the jet,
$p \to \infty$, where momentum broadening is dominated by small-angle
$t$-channel scattering processes \cite{Arnold:2000dr,Boguslavski:2023waw}.
We begin with the spacelike four-momentum transfer,
\begin{equation}
Q^2 = \mathbf{q}^{2} - \omega^2 \ge 0.
\end{equation}
The integration over
$\mathbf{p}'$ then follows from the spatial delta function, yielding
\begin{equation}
\hat q^{ij}_{\mathrm{eq}}
=
\frac{1}{4 p\, \nu_a}
\sum_{bd}
\int d\Gamma_{k k'}
q^i q^j \,(2\pi)\delta(p+k-p'-k')
\left|\mathcal{M}^{ab}_{cd}\right|^2 f^b_{0k}\tilde f^d_{0k'}.
\end{equation}
For the remaining integrals, we first introduce an integral over the energy
transfer $\omega$,
\begin{equation}
\delta(p+k-p'-k')
= \int_{-\infty}^{\infty} d\omega\,
\delta(\omega + p - p')\,\delta(\omega - k + k'),
\end{equation}
and rewrite the delta functions in terms of angular variables,
\begin{align}
\delta(\omega + p - p')
&= \frac{1}{2pq}
\delta\!\left(\cos\theta_{pq}-\frac{\omega}{q}-\frac{t}{2pq}\right)
\Theta(\omega+p), \\
\delta(\omega - k + k')
&= \frac{1}{2kq}
\delta\!\left(\cos\theta_{kq}-\frac{\omega}{q}+\frac{t}{2kq}\right)
\Theta(k-\omega),
\end{align}
where $t = \omega^2 - \mathbf{q}^{\,2} \leq 0$ is the spacelike Mandelstam
variable. This fixes the angular integrations, which can then be performed
trivially. After the change of variables $\mathbf{k}'=\mathbf{k}-\mathbf{q}$,
one obtains
\begin{align}
\hat q^{ij}_{\mathrm{eq}}
=&
\frac{1}{4 p^2\, \nu_a (2\pi)^5}
\sum_{bcd}
\int_0^\infty dq \int_{-q}^q d\omega
\int_{\frac{q+\omega}{2}}^{p+2\omega} dk
\int_0^{2\pi} d\phi_{pq}
\int_0^{2\pi} d\phi_{kq}\,
q^i q^j \left|\mathcal{M}^{ab}_{cd}\right|^2  f^b_{0k}\tilde f^d_{0k'}.
\end{align}
We note that, in the limit where $p\rightarrow \infty$, the upper bound of the integration over $k$ can be taken to be infinity. In an isotropic medium, the momentum transfer is parametrized in terms of the
azimuthal and polar angles between the $q$-$p$ and $q$-$k$ directions such
that, after averaging over azimuthal angles, the tensor structure of
$\hat q_{\mathrm{eq}}^{ij}$ is preserved. We use the decomposition
\begin{align}
Q^0 &= \omega, \\
Q^1 &= q\sin\theta_{pq}\cos\phi_{pq}, \\
Q^2 &= q\sin\theta_{pq}\sin\phi_{pq}, \\
Q^3 &= q\cos\theta_{pq}.
\end{align}
The tensor structure follows from rotational symmetry around the jet axis and is
independent of the detailed form of the matrix element. In particular, integrals
over odd functions of $\cos\phi_{pq}$ or $\sin\phi_{pq}$ vanish, simplifying the
structure of $\hat q_{\mathrm{eq}}^{ij}$. Since the integrand is invariant under
$\phi_{pq}\to\phi_{pq}+\pi$, all contributions odd under this transformation
vanish. In particular,
\begin{equation}
\int_0^{2\pi} d\phi_{pq}\,\cos\phi_{pq} =
\int_0^{2\pi} d\phi_{pq}\,\sin\phi_{pq} = 0,
\end{equation}
and more generally, any term containing an odd power of $\cos\phi_{pq}$ or
$\sin\phi_{pq}$ integrates to zero. It follows that all components linear in
$q^1$ or $q^2$ vanish after angular integration, yielding
\begin{equation}
\hat q^{12} = \hat q^{13} = \hat q^{23} = 0.
\end{equation}
The equality $\hat q^{11}=\hat q^{22}\equiv\hat q_\perp$ is a direct consequence
of the isotropy of the medium, since the two transverse directions are equivalent
for the jet. The remaining diagonal components may be nonzero, resulting in the
tensor structure
\begin{equation}
\hat q^{ij}_{\mathrm{eq}} =
\begin{pmatrix}
\hat q^{11} & 0 & 0 \\
0 & \hat q^{22} & 0 \\
0 & 0 & \hat q^{33}
\end{pmatrix},
\qquad
\hat q^{11}=\hat q^{22}= \hat q_\perp\, , \qquad \hat q^{33} = \hat q_\parallel.
\end{equation}

We now turn to the matrix element. For clarity, we focus on a jet propagating
through a purely gluonic medium with all possible $2\leftrightarrow 2$ scattering
processes; the inclusion of quarks proceeds analogously, with the appropriate
replacement of color factors and statistical functions, as noted in the main text.
In the high-energy limit, the matrix element is written in terms of a screened interaction based on the HTL form \cite{Arnold:2003zc,Kurkela:2018oqw,Schlichting:2020lef},
following the $\xi$-screening prescription, in which $\xi = e^{1/3}/2$ is the infrared screening parameter \cite{AbraaoYork:2014hbk},
which has been shown to be in numerical agreement with the conditions considered
here \cite{Boguslavski:2024jwr}. Explicitly,
\begin{equation}
\tilde{\mathcal{M}}_{\xi}^2
= \lim_{p\to\infty} \frac{\left|\mathcal{M}^{ab}_{cd}\right|^2}{g^4p^2}
= \frac{4\left( 2k - \omega - \sqrt{(2k - \omega)^2 - q^2}\cos\phi_{kq} \right)^2}
{16\left( q^2 + \xi^2 m_D^2 \right)^2},
\end{equation}
where the factor $16 = 2^4$ in the denominator accounts for the relativistic
phase-space normalization $2E$ of each external particle, absorbed into
$\tilde{\mathcal{M}}_{\xi}^2$ for consistency with the Lorentz-invariant
phase-space measure $d\Gamma = \frac{d^3\mathbf{k}}{(2\pi)^3 k^0}$, and
$m_D^2 = N_c g^2 T^2/3 + N_F g^2(T^2 + 3\mu^2/\pi^2)/6$ is the Debye mass
squared \cite{Rischke:2003mt}. We note that the general form of the diagram is accompanied with the appropriate group factors for gluon or quark scattering. Schematically,
\begin{equation}
\hat q^{ij}_{\mathrm{eq}}
\sim
\int_0^\infty dq \int_{-q}^q d\omega
\int_{\frac{q+\omega}{2}}^{p+2\omega} dk \,
q^i q^j \,
\tilde{\mathcal{M}}_{\xi}^2 \,
f^b_{0k}\tilde f^d_{0k'}.
\end{equation}
For the analytical computation in equilibrium, one can further approximate
$\tilde{f}^d_{0k'} \approx \tilde{f}^d_{0k}$, since in the soft limit
$\omega \ll k$ the outgoing momentum $k' = k - \omega \approx k$. Under this
approximation, the statistical factor reduces to
$f^b_{0k}\tilde{f}^b_{0k} = f^b_{0k}(1+f^b_{0k})$,
and the $k$ integral can be performed analytically using
\begin{equation}
\int_0^\infty dk\, k^2\, f_{0k}(1+f_{0k}) = \frac{\pi^2 T^3}{6},
\end{equation}
yielding the results quoted in Eqs.~(\ref{eq:qhat_eq}) and~(\ref{eq:qhatL_eq}).
For the fully equilibrated medium, the standard transverse and longitudinal
broadening coefficients are
\begin{align}
\hat{q} &\equiv \hat{q}^{11}_{\rm eq} =
\frac{g^4 C_A}{4\pi} T m_D^2
\ln\!\left(1+\frac{\Lambda_\perp^2}{m_D^2}\right),
\label{eq:qhat_eq}\\
\hat{q}_L &\equiv \hat{q}^{33}_{\rm eq} =
\frac{g^2 C_A T\, M_\infty^2}{4\pi}
\ln\!\left(1+\frac{\mu_{\tilde q_\perp}^2}{M_\infty^2}\right),
\label{eq:qhatL_eq}
\end{align}
where $m_D$ is the Debye mass,
\begin{equation}
M_\infty^2 = \frac{g^2 C_A T^2}{6}
\end{equation}
is the gluon asymptotic thermal mass, $\mu_{\tilde q_\perp}$ denotes the
transverse-momentum matching scale separating soft and hard contributions to
the longitudinal broadening coefficient \cite{Ghiglieri:2015ala}, $C_A$
is the Casimir of the adjoint representation of $SU(3)$ and $\Lambda_\perp$ denotes the upper bound for the momentum exchanged between the jet and the medium parton.

The equilibrium results presented here serve two purposes: they confirm that
our formulation reproduces the tensor structure of $\hat q^{ij}$ found in the
literature, and they provide the baseline against which the non-equilibrium
corrections discussed in the main text are measured. 

\section{Jet momentum broadening for an anisotropic medium}

We now compute the leading non-equilibrium correction to the jet momentum-broadening
tensor, $\delta\hat{q}^{ij}$, from Eq.~(\ref{eq:define_delta_qij}). Under the
assumptions discussed above, the medium anisotropy enters solely through corrections
to the thermal distribution. These corrections break the isotropy of the medium and
introduce flow effects, so that the symmetries exploited in the previous section no
longer apply. To systematically incorporate these corrections, following
\cite{Denicol:2012cn}, we expand $\phi_k$ in a complete and orthogonal basis of
irreducible tensors constructed from $K^\mu$, $U^\mu$, and the metric $g_{\mu\nu}$,
\begin{equation}
1,\; K^{\langle\mu\rangle},\; K^{\langle\mu}K^{\nu\rangle},\;
K^{\langle\mu}K^{\nu}K^{\lambda\rangle},\ldots\,.
\end{equation}
Angle brackets denote symmetric, traceless projections orthogonal to the fluid
four-velocity,
\begin{equation}
A^{\langle\mu\rangle} = \Delta^\mu{}_\nu A^\nu,
\qquad
\Delta_{\mu\nu} = g_{\mu\nu} + U_\mu U_\nu,
\end{equation}
with higher-rank projections defined analogously. The expansion then reads
\cite{Denicol:2012cn}
\begin{align}
\phi_k
=
\sum_{\ell=0}^{\infty}\sum_{n=0}^{N_\ell}
\mathcal{H}_{\mathbf{k}n}^{(\ell)}
\rho_n^{\mu_1\cdots\mu_\ell}
K_{\langle\mu_1}\cdots K_{\mu_\ell\rangle},
\end{align}
where the moments are defined as
\begin{equation}
\rho_n^{\mu_1\cdots\mu_\ell}(X) = \int \frac{d^3\mathbf{k}}{(2\pi)^3 k^0}
E_{\mathbf{k}}^n\, K^{\langle\mu_1}\cdots K^{\mu_\ell\rangle}\,\delta f(X,K),
\end{equation}
with $\mathcal{H}_{\mathbf{k}n}^{(\ell)}$ being polynomials that depend only on
$E_{\mathbf{k}}$, and the integer $\ell$ labeling the rank of the irreducible
tensors \cite{Denicol:2012cn}. Typical Landau matching conditions \cite{Rocha:2023ilf}
correspond to setting $\rho_1 = \rho_2 = \rho_1^\mu = 0$, and the shear-stress
tensor is identified as $\pi^{\mu\nu} = \rho_0^{\mu\nu}$ \cite{Denicol:2012cn}.

Exact equations of motion for the moments $\rho_n^{\mu_1\cdots\mu_\ell}$ can be
obtained from the underlying integro-differential Boltzmann equation
\cite{Denicol:2012cn}. In principle, solving this infinite hierarchy of coupled
partial differential equations is equivalent to solving the Boltzmann equation
itself and can be used to reconstruct the full distribution function; see
\cite{Bazow:2015dha,Bazow:2016oky,Mullins:2022fbx} for explicit examples in
relativistic systems. For the present work, however,
Eq.~\eqref{eq:define_delta_qij} shows that any approximation for $\phi_k$
obtained either from the moment hierarchy itself or from a physically motivated
truncation scheme yields an approximation to the out-of-equilibrium contribution
$\delta\hat{q}^{ij}$ generated by the medium. For simplicity, we adopt the
lowest-order truncation compatible with causal and stable hydrodynamic evolution
\cite{Rocha:2023ilf}, namely the fourteen-moment approximation
\cite{IsraelStewart1979,UdeyIsrael1982}, in which the series is truncated at
rank two so that moments with $\ell \ge 3$ are neglected,
\begin{equation}
\rho_r^{\mu_1\cdots\mu_\ell} = 0, \qquad \ell \ge 3.
\end{equation}
This retains only the scalar ($\ell=0$), vector ($\ell=1$), and tensor ($\ell=2$)
moments, corresponding respectively to bulk viscous pressure $\Pi$, baryon
diffusion current $n^\mu$, and shear-stress tensor $\pi^{\mu\nu}$. The resulting
expression for $\delta\hat{q}^{ij}$ can be systematically improved by including
higher-rank moments \cite{Denicol:2012cn}. In general, one may write
\cite{Denicol:2012cn}
\begin{align}
\phi_k
=&\,
\bigl[E_0 + B_0 m^2 - D_0 E_k - 4B_0 E_k^2\bigr]\Pi
+ \lambda_n n_\alpha K^\alpha
+ B_2\,\pi_{\alpha\beta} K^\alpha K^\beta\,.
\label{eq:matching}
\end{align}
For the massless system considered here, bulk viscosity and diffusion are neglected,
so that $\Pi = 0$ and $n^\mu = 0$. The only surviving dissipative structure is
therefore the rank-two moment $\pi^{\mu\nu} = \rho_0^{\mu\nu}$, and accordingly
\begin{equation}
\phi_k = B_2\,\pi_{\alpha\beta}K^\alpha K^\beta,
\qquad B_2 = \frac{1}{2J_{42}},
\label{eq:phi_shear_general_clean}
\end{equation}
where
\begin{equation}
J_{nl} \equiv \int \frac{d^3\mathbf{k}}{(2\pi)^3\,E_{\mathbf{k}}}\,
E_{\mathbf{k}}^{\,n-2l}\,
\bigl(\Delta_{\alpha\beta} K^\alpha K^\beta\bigr)^{l}\,
f_0(k)\,.
\end{equation}
An analogous expression holds for $\phi_{k'}$, such that in the following
derivations we present the details as a function of $K$ and then assume that
the same procedure is applied to $K'$.

\subsection*{Linearized correction to $\hat q^{ij}$}
Keeping only terms linear in $\delta f$, the non-equilibrium correction to the 
broadening tensor is
\begin{align}
\delta\hat q^{ij}
=&\frac{1}{4\nu_a p}
\sum_{bd}
\int d\Gamma_{k p' k'}
\,
q^i q^j
(2\pi)^4\delta^{(4)}(P+K-P'-K')
|\mathcal{M}^{ab}_{cd}(P,K;P',K')|^2 
\nonumber\\
&\times
\Big[
f_{0k}f_{0k'}\tilde f_{0k'}\,\phi_{k'}
+
f_{0k}\tilde f_{0k}\tilde f_{0k'}\,\phi_k
\Big].
\label{eq:dqij_start_clean}
\end{align}
Substituting Eq.~\eqref{eq:phi_shear_general_clean}, one finds
\begin{align}
\delta\hat q^{ij}
&=\,\frac{1}{4\nu_a (2\pi)^5}
\sum_{bd}
\int d\Gamma_{k p' k'}
\,
q^i q^j (2\pi)\delta(p+k-p'-k')\,
|\mathcal{M}^{ab}_{cd}(P,K;P',K')|^2
f_{0k}\tilde f_{0k'}
\nonumber\\
&\times
\left[
\frac{f_{0k'}}{2J_{42}}\,
\pi_{\mu\nu}K'^\mu K'^\nu
+
\frac{\tilde f_{0k}}{2J_{42}}\,
\pi_{\mu\nu}K^\mu K^\nu
\right].
\label{eq:dqij_shear_clean}
\end{align}
We emphasize that the formation of Eq.~(\ref{eq:dqij_shear_clean}) is completely general, and no assumptions about the particular form of the interaction considered are necessary up to this point. Furthermore, the shear tensor used in this calculation is an input that is taken from hydrodynamical description of the medium, and has no explicit connection to the interaction considered here.

\subsection*{Local rest frame}
For simplicity, we perform this integration in the local rest frame, $U^\mu=(1,\mathbf{0})$, the shear tensor is purely spatial,
\begin{equation}
\pi^{0\mu}=0,
\qquad
\pi^{\mu\nu}\rightarrow \pi^{ij},
\end{equation}
so that
\begin{equation}
\phi_k = \frac{1}{2J_{42}}\,\pi_{ij}k^i k^j.
\end{equation}
Thus the entire non-equilibrium correction is controlled by the anisotropic contraction
$\pi_{ij}k^i k^j$. The evaluation of $\delta\hat{q}^{ij}$ proceeds in exact analogy with the
equilibrium calculation: we first perform the
$\mathbf{p}'$ integration using the spatial delta function, then introduce
the auxiliary integration over the energy transfer $\omega$, with
$t = \omega^2 - q^2$. Following the same steps as before, one arrives at,
\begin{align}
\delta\hat q^{ij}
=&\sum_{a}\,\frac{1}{4\nu_a(2\pi)^5}
\int_0^\infty dq
\int_{-q}^{q}d\omega
\int_{\frac{q+\omega}{2}}^{p+2\omega}dk
\int_0^{2\pi}d\phi_{pq}\int_0^{2\pi}d\phi_{kq}
\int_{-1}^{1}d\cos\theta_{kq}\int_{-1}^{1}d\cos\theta_{pq}\;
q^i q^j
\nonumber\\
&\times \tilde{\mathcal{M}}_{\xi}^2\,
\delta\!\left(
\cos\theta_{pq}-\frac{\omega}{q}-\frac{t}{2pq}
\right)\,
\delta\!\left(
\cos\theta_{kq}+\frac{\omega}{q}-\frac{t}{2kq}
\right)
f_{0k}\tilde f_{0k'}
\left[
\frac{f_{0k'}}{2J_{42}}\pi_{mn}k'^m k'^n
+
\frac{\tilde f_{0k}}{2J_{42}}\pi_{mn}k^m k^n
\right].
\label{eq:dqij}
\end{align}

\subsection*{Explicit non-diagonal structure}
We now discuss the tensor structure of $\delta\hat{q}^{ij}$ and the choice of
basis in which the calculation is most naturally performed. Since the medium is
no longer isotropic, the simplifications exploited in the equilibrium calculation
no longer apply, and the full tensor structure must be retained. A key point is
that $\pi^{\mu\nu}$ is naturally defined in the jet-fixed frame, whereas the
delta function fixing $\theta_{kq}$ is most naturally expressed in a basis
aligned with the momentum transfer $\mathbf{q}$. The non-equilibrium angular
structure is therefore not captured correctly unless $k^i$ is rotated back into
the jet-fixed basis before evaluating the contraction $\pi_{\alpha\beta}
K^\alpha K^\beta$. We now make this change of basis explicit.

We choose the jet direction as $\hat{n}$ and introduce an orthonormal basis
$\{\hat{e}_1,\hat{e}_2,\hat{e}_3\}$ adapted to the momentum-transfer direction
$\hat{e}_3 = \mathbf{q}/q$,
\begin{equation}
\hat{e}_3
=
\big(
\sqrt{1-\eta^2}\cos\phi_{pq},\,
\sqrt{1-\eta^2}\sin\phi_{pq},\,
\eta
\big),
\qquad
\eta \equiv \cos\theta_{pq}
=
\frac{\omega}{q}+\frac{t}{2pq}\xrightarrow{p\to\infty}\frac{\omega}{q},
\end{equation}
where the last step uses $t/2pq\ll\omega/q$ in the high-momentum probe limit, and
\begin{equation}
\hat{e}_1= \big( \eta\cos\phi_{pq},\, \eta\sin\phi_{pq},\, -\sqrt{1-\eta^2} \big),
\qquad
\hat{e}_2= \big( -\sin\phi_{pq},\, \cos\phi_{pq},\, 0 \big).
\label{eq:basis}
\end{equation}
In this basis, the medium momentum $\mathbf{k}$ is decomposed as
\begin{equation}
\mathbf{k}
=
k\Big[
\sqrt{1-\rho^2}\cos\phi_{kq}\,\hat{e}_1
+
\sqrt{1-\rho^2}\sin\phi_{kq}\,\hat{e}_2
+
\rho\,\hat{e}_3
\Big],
\qquad
\rho \equiv \cos\theta_{kq}
=
\frac{\omega}{q}-\frac{t}{2kq}.
\end{equation}
The delta functions can then be used to fix $\rho = \cos\theta_{kq}$ and
$\eta = \cos\theta_{pq}$, reducing the angular integrations to the two
remaining azimuthal angles $\phi_{kq}$ and $\phi_{pq}$. However, the tensor $k^i k^j$ expressed in the jet-fixed basis depends on both

\subsection*{Explicit harmonic projection and emergence of mixed components}

Before performing the final integrals, we evaluate explicitly the components of
the tensor $k^ik^j$ in the jet-fixed frame. This intermediate step allows one to
decompose the integrand in powers of $\cos\phi_{pq}$ and $\cos\phi_{kq}$, so that
the angular integrals can be performed directly and the vanishing contributions
identified by inspection. The anisotropic structure of $\delta\hat{q}^{ij}$ can
therefore be understood by tracking the angular harmonics of the integrand in the
jet frame.

Consider in particular the contribution proportional to $\pi^{xz}$. From the
rotated basis defined in Eq.~(\ref{eq:basis}), the $x$-component of $\mathbf{k}$
in the jet frame is
\begin{equation}
k^x =
k\Big[
\sqrt{1-\rho^2}\cos\phi_{kq}\,e_1^x
+
\sqrt{1-\rho^2}\sin\phi_{kq}\,e_2^x
+
\rho\,e_3^x
\Big],
\end{equation}
where the explicit basis-vector components are
\begin{equation}
e_1^x = \eta\cos\phi_{pq},
\qquad
e_2^x = -\sin\phi_{pq},
\qquad
e_3^x = \sqrt{1-\eta^2}\cos\phi_{pq}.
\end{equation}
Substituting these, one obtains
\begin{align}
k^x =
k\Big[
\sqrt{1-\rho^2}\cos\phi_{kq}\,\eta\cos\phi_{pq}
-\sqrt{1-\rho^2}\sin\phi_{kq}\,\sin\phi_{pq}
+
\rho\sqrt{1-\eta^2}\cos\phi_{pq}
\Big].
\end{align}
Similarly, the $z$-component is
\begin{equation}
k^z =
k\Big[
-\sqrt{1-\rho^2}\cos\phi_{kq}\sqrt{1-\eta^2}
+
\rho\eta
\Big].
\end{equation}
The contraction $\pi_{\alpha\beta}K^\alpha K^\beta$ contains the term
$2\pi^{xz}k^xk^z$, where $\pi^{xz}$ is a numerical input from hydrodynamics
and introduces no additional angular structure. Substituting the expressions
above, one finds
\begin{align}
k^xk^z
&=
k^2
\Big[
-(1-\rho^2)\eta\sqrt{1-\eta^2}\cos^2\phi_{kq}\cos\phi_{pq}
+(1-\rho^2)\sqrt{1-\eta^2}\sin\phi_{kq}\cos\phi_{kq}\sin\phi_{pq}
\nonumber\\
&\qquad\quad
-\rho\eta\sqrt{1-\rho^2}\sin\phi_{kq}\sin\phi_{pq}
+\rho\sqrt{1-\rho^2}(2\eta^2-1)\cos\phi_{kq}\cos\phi_{pq}
\nonumber\\
&\qquad\quad
+\rho^2\eta\sqrt{1-\eta^2}\cos\phi_{pq}
\Big].
\end{align}
The remaining components are
\begin{align}
k^xk^x
&=
k^2\Big[
(1-\rho^2)\eta^2\cos^2\phi_{kq}\cos^2\phi_{pq}
+(1-\rho^2)\sin^2\phi_{kq}\sin^2\phi_{pq}
+\rho^2(1-\eta^2)\cos^2\phi_{pq}
\nonumber\\
&\qquad\quad
-2(1-\rho^2)\eta\sin\phi_{kq}\cos\phi_{kq}\sin\phi_{pq}\cos\phi_{pq}
+2\rho\eta\sqrt{1-\rho^2}\sqrt{1-\eta^2}\cos\phi_{kq}\cos^2\phi_{pq}
\nonumber\\
&\qquad\quad
-2\rho\sqrt{1-\rho^2}\sqrt{1-\eta^2}\sin\phi_{kq}\sin\phi_{pq}\cos\phi_{pq}
\Big],
\\[1ex]
k^yk^y
&=
k^2\Big[
(1-\rho^2)\eta^2\cos^2\phi_{kq}\sin^2\phi_{pq}
+(1-\rho^2)\sin^2\phi_{kq}\cos^2\phi_{pq}
+\rho^2(1-\eta^2)\sin^2\phi_{pq}
\nonumber\\
&\qquad\quad
+2(1-\rho^2)\eta\sin\phi_{kq}\cos\phi_{kq}\sin\phi_{pq}\cos\phi_{pq}
+2\rho\eta\sqrt{1-\rho^2}\sqrt{1-\eta^2}\cos\phi_{kq}\sin^2\phi_{pq}
\nonumber\\
&\qquad\quad
+2\rho\sqrt{1-\rho^2}\sqrt{1-\eta^2}\sin\phi_{kq}\sin\phi_{pq}\cos\phi_{pq}
\Big],
\\[1ex]
k^zk^z
&=
k^2\Big[
(1-\rho^2)(1-\eta^2)\cos^2\phi_{kq}
-2\rho\eta\sqrt{1-\rho^2}\sqrt{1-\eta^2}\cos\phi_{kq}
+\rho^2\eta^2
\Big],
\\[1ex]
k^xk^y
&=
k^2\Big[
(1-\rho^2)\eta^2\cos^2\phi_{kq}\sin\phi_{pq}\cos\phi_{pq}
-(1-\rho^2)\sin^2\phi_{kq}\sin\phi_{pq}\cos\phi_{pq}
+\rho^2(1-\eta^2)\sin\phi_{pq}\cos\phi_{pq}
\nonumber\\
&\qquad\quad
+(1-\rho^2)\eta\sin\phi_{kq}\cos\phi_{kq}\big(\cos^2\phi_{pq}-\sin^2\phi_{pq}\big)
+2\rho\eta\sqrt{1-\rho^2}\sqrt{1-\eta^2}\cos\phi_{kq}\sin\phi_{pq}\cos\phi_{pq}
\nonumber\\
&\qquad\quad
+\rho\sqrt{1-\rho^2}\sqrt{1-\eta^2}\sin\phi_{kq}\big(\cos^2\phi_{pq}-\sin^2\phi_{pq}\big)
\Big],
\\[1ex]
k^yk^z
&=
k^2\Big[
-(1-\rho^2)\eta\sqrt{1-\eta^2}\cos^2\phi_{kq}\sin\phi_{pq}
-(1-\rho^2)\sqrt{1-\eta^2}\sin\phi_{kq}\cos\phi_{kq}\cos\phi_{pq}
\nonumber\\
&\qquad\quad
+\rho\eta\sqrt{1-\rho^2}\sin\phi_{kq}\cos\phi_{pq}
+\rho\sqrt{1-\rho^2}(2\eta^2-1)\cos\phi_{kq}\sin\phi_{pq}
\nonumber\\
&\qquad\quad
+\rho^2\eta\sqrt{1-\eta^2}\sin\phi_{pq}
\Big].
\end{align}
The tensor $q^iq^j$ likewise contains harmonics in $\phi_{pq}$; for example,
\begin{equation}
q^xq^z = q^2\,\eta\sqrt{1-\eta^2}\cos\phi_{pq}.
\end{equation}
The angular integrals therefore receive nonzero contributions only from terms
with even powers of $\cos\phi_{kq}$, while terms with odd powers integrate to
zero. The selection rules for $\phi_{pq}$ are determined by the harmonics in
$q^iq^j$, and all contributions must be accounted for before proceeding to the
remaining integrals. For the component considered here,
\begin{align}
\delta\hat{q}^{xz} \sim \int q^xq^z\,\pi_{mn}k^m k^n\,
\tilde{\mathcal{M}}_\xi^2\,f^b_{0k}\tilde{f}^d_{0k'}.
\end{align}
After integrating over $\phi_{pq}$ and $\phi_{kq}$, one obtains
\begin{equation}
\begin{aligned}
\delta\hat{q}^{xz} &= -\frac{g^4C_A}{\nu_a(2\pi)^5}
\frac{\pi^2 \, k^2 \omega q}{4\left( q^2 + \xi^2 m_D^2 \right)^2}
\Bigg\{
8(2k-\omega)\sqrt{(2k-\omega)^2-q^2}\left[7(2k-\omega)^2-3q^2\right]
\left(1-\frac{2\omega^2}{q^2}\right)\eta\sqrt{\frac{t(\eta^2-1)}{q^2}}
\\[6pt]
&+ \frac{\omega(\omega^2-q^2)}{q^3}
\Big[
(119\eta^2-49)(2k-\omega)^4
+(46-106\eta^2)(2k-\omega)^2 q^2
+(11\eta^2-5)q^4
\Big]\\
&\times \frac{f^b_{0k}\tilde{f}^b_{0k}\tilde{f}^d_{0k'}
+f^b_{0k}\tilde{f}^d_{0k'}f^d_{0k'}}{2J_{42}}
\Bigg\} \times \pi^{xz},
\end{aligned}
\end{equation}
where the remaining integrals over $q$, $\omega$, and $k$ are performed
numerically. This demonstrates explicitly that the mixed component
$\delta\hat{q}^{xz}$ survives the angular integrations. Note that the argument of the square root is non-negative: since $t = \omega^2 - q^2 \leq 0$
and $\eta^2 \leq 1$, one has $t(\eta^2-1) \geq 0$, so the expression is real throughout
the integration domain. The same mechanism
applies to all components: the rotated basis introduces $\phi_{pq}$-dependent
harmonics into $k^i$, which combine with those from $q^iq^j$ to produce
non-vanishing projections onto $\pi^{xy}$, $\pi^{xz}$, and $\pi^{yz}$. If the
rotation is neglected and $k^i$ is treated as depending only on $\phi_{kq}$,
these mixed harmonics are absent and the corresponding anisotropic contributions
are incorrectly suppressed.

The remaining integrations over $q$, $\omega$, and $k$ are performed numerically
using Gaussian quadrature. In what follows, we analyze the tensorial structure of
$\delta\hat{q}^{ij}$ in more detail and derive its most general decomposition in
terms of the available symmetric tensors constructed from the fluid velocity
$U^\mu$, the jet direction $\hat{n}$, and the shear-stress tensor $\pi^{\mu\nu}$.

\subsection*{General tensorial structure of $\delta\hat{q}^{ij}$}

The explicit angular analysis shows that $\delta\hat{q}^{ij}$ does not vanish
for the mixed components. Rather, the rotation between the $\mathbf{q}$- and
$\mathbf{k}$-aligned frames generates nontrivial angular correlations that
survive the azimuthal integrations and produce finite projections onto the
off-diagonal components of $\pi^{\mu\nu}$.

After performing the angular integrations, the result must reduce to a symmetric
rank-two tensor constructed from the jet direction $\hat{n} \equiv \hat{z}$ and
the shear-stress tensor $\pi^{ij}$. The most general form consistent with these
symmetries is
\begin{equation}
\delta \hat{q}^{ij}
=
\alpha\,\pi^{ij}
+\beta\,\delta^{ij}\pi^{zz}
+\gamma\left(
\hat{z}^i\pi^{jz}
+\hat{z}^j\pi^{iz}
+\hat{z}^i\hat{z}^j\pi^{zz}
\right),
\label{eq:dqij_decomp}
\end{equation}
where the coefficients $\alpha$, $\beta$, and $\gamma$ are determined numerically
and depend on the transverse momentum cutoff $\Lambda_\perp$ and the screening
mass $m_D$. As the momentum exchanged between the jet and the medium partons is integrated over all possible values, the integral from $0$ to $\infty$ diverges, indicating that the high-momentum-exchange limit is not well suited to describe the system. Furthermore, the perturbative framework presented here is valid only for small momentum exchange; therefore, we set the maximum momentum exchange to $5T$, which lies within the perturbative regime while remaining close to the mean thermal momentum of medium constituents, of order $3T$ in QCD kinetic theory for a massless gas, as shown in the figure in the main text.

In this basis, the individual components take the explicit matrix form
\begin{equation}
\delta \hat{q}^{ij}
=
\begin{pmatrix}
\alpha\pi^{xx}+\beta\pi^{zz} & \alpha\pi^{xy} & (\alpha+\gamma)\pi^{xz} \\
\alpha\pi^{xy} & \alpha\pi^{yy}+\beta\pi^{zz} & (\alpha+\gamma)\pi^{yz} \\
(\alpha+\gamma)\pi^{xz} & (\alpha+\gamma)\pi^{yz} & (\alpha+\beta+3\gamma)\pi^{zz}
\end{pmatrix}.
\end{equation}

At the integrand level, each component $\delta\hat{q}^{ij}$ receives contributions
from the full contraction $\mathcal{K}^{ijmn}\pi_{mn}$. The coefficients $\alpha$,
$\beta$, and $\gamma$ are obtained by performing the angular integrations and
projecting the resulting rank-four tensor onto the symmetry-allowed basis.
Using the notation of Eq.~\eqref{eq:define_delta_qij} and the matrix element
$\tilde{\mathcal{M}}_\xi^2$ defined above, they read explicitly
\begin{align}
\alpha
=&\,\frac{1}{4\nu_a(2\pi)^5}
\int d\Gamma\;
q^x q^y\,\tilde{\mathcal{M}}_{\xi}^2\,
f^b_{0k}\tilde{f}^d_{0k'}
\left[
\frac{f^d_{0k'}}{2J_{42}}\,k'^xk'^y
+
\frac{\tilde{f}^b_{0k}}{2J_{42}}\,k^xk^y
\right],
\label{eq:alpha}
\\[1ex]
\beta
=&\,\frac{1}{4\nu_a(2\pi)^5}
\int d\Gamma\;
q^x q^x\,\tilde{\mathcal{M}}_{\xi}^2\,
f^b_{0k}\tilde{f}^d_{0k'}
\left[
\frac{f^d_{0k'}}{2J_{42}}\,k'^zk'^z
+
\frac{\tilde{f}^b_{0k}}{2J_{42}}\,k^zk^z
\right],
\label{eq:beta}
\\[1ex]
\gamma
=&\,\frac{1}{4\nu_a(2\pi)^5}
\int d\Gamma\;
q^x q^z\,\tilde{\mathcal{M}}_{\xi}^2\,
f^b_{0k}\tilde{f}^d_{0k'}
\left[
\frac{f^d_{0k'}}{2J_{42}}\,k'^xk'^z
+
\frac{\tilde{f}^b_{0k}}{2J_{42}}\,k^xk^z
\right]
-\alpha.
\label{eq:gamma}
\end{align}
This shows that the off-diagonal components of $\delta\hat{q}^{ij}$ are directly
controlled by the corresponding components of $\pi^{\mu\nu}$, and vanish only in
the principal-axis frame of $\pi^{ij}$ where the shear tensor is diagonal.
Therefore, once the full angular structure induced by the rotated basis is
retained, shear flow generically generates nonzero off-diagonal contributions to
the broadening tensor, underscoring the importance of retaining the full tensorial
structure of $\delta\hat{q}^{ij}$ for a consistent description of jet broadening
out of equilibrium, as emphasized in the main text.

We also note that a fully covariant extension of the momentum-broadening tensor
can be constructed for QCD as well as for scalar theories following the approach
of Ref.~\cite{Danhoni:2026gve}. In such a formulation, additional components
arise beyond the transverse sector considered here; in particular, azimuthal
symmetry does not restrict components involving energy-momentum correlations,
so $\hat q^{0\nu}$ need not vanish. We leave this covariant extension for
future work.


\end{document}